\documentclass[useAMS,usenatbib]{mn2e}

\usepackage{amssymb}
\usepackage[dvips]{graphicx,color}
\usepackage{natbib}
\usepackage{multirow}
\title[Asteroseismology to investigate dark matter]{Towards the use of asteroseismology to investigate the nature of dark matter}
\author[Jordi Casanellas and Il\'idio Lopes]{Jordi Casanellas$^{1}$\thanks{E-mail:
jordicasanellas@ist.utl.pt} and Il\'idio Lopes$^{1,2}$\thanks{E-mail: ilidio.lopes@ist.utl.pt}\\
$^{1}$Centro Multidisciplinar de Astrof\'isica, Instituto Superior T\'ecnico, Av. Rovisco Pais, 1049-001 Lisboa, Portugal\\
$^{2}$Departamento de F\'isica, Universidade de \'Evora, Col\'egio Luis Ant\'onio Verney, 7002-554 \'Evora, Portugal}
\begin{document}

\date{}

\pagerange{\pageref{firstpage}--\pageref{lastpage}} \pubyear{2002}

\maketitle

\label{firstpage}

\begin{abstract}
The annihilation of huge quantities of captured dark matter (DM) particles inside low-mass stars has been shown to change some of the stellar properties, such as the star's effective temperature or the way the energy is transported throughout the star. While in the classical picture, without DM, a star of 1 M$_{\odot}$ is expected to have a radiative interior during the main sequence, the same star evolving in a halo of DM with a density $\rho_{\chi}>10^{8}\;$GeV$\;$cm$^{-3}$ will develop  a convective core in order to evacuate the energy from DM annihilation in a more efficient way. This convective core leaves a discontinuity in the density and sound-speed profiles that can be detected by the analysis of the stellar oscillations. In this paper we present an approach towards the use of asteroseismology to detect the signature produced by the presence of DM inside a star, and we propose a new methodology to infer the properties of a DM halo from the stellar oscillations (such as the product of the DM density and the DM particle-nucleon scattering cross-section).
\end{abstract}

\begin{keywords}
dark matter - asteroseismology - stars: interiors - stars: low-mass - stars: fundamental parameters - Galaxy: centre
\end{keywords}

\section{Introduction}
Different observations in a wide range of scales, from galactic to cosmological, suggest the existence of a new kind of matter, called Dark Matter (DM), formed by unknown particles. Among the possible constituents of DM, the WIMPs, massive particles with non-negligible scattering cross-section with baryons, are considered one of the best candidates (\citealt*{rev-BertoneHS2005}).

Soon was realised that, if WIMPs exist, they will accumulate inside stars \citep{art-PressSpergel1985} and their annihilation may lead to significant changes in the classical picture of stellar evolution if the halo where the stars evolve has a very high density of DM particles \citep{art-BouquetSalati1989,art-SalatiSilk1989,art-Dearbornetal1990}. In this context, the effects of the capture of WIMPs by the Sun were studied, addressing the prospects of helioseismology to test models that solved the old solar neutrino problem \citep{art-DappenGilliandCh-D1986,art-FaulknerGoughVahia1986}, and to give constraints to the nature of DM particles (\citealt*{art-LopesSH2002,let-LopesSilk2002,art-LopesBS2002}; \citealt{art-Bottinoetal2002}; \citealt{art-Cumberbatchetal2010}; \citealt{art-FrandsenSarkar2010PhRvL}; \citealt{art-Taosoetal2010}).

Recently, particular attention has been given to the first stars formed in the early Universe due to the high DM content in that epoch (\citealt*{let-Spolyaretal2008}; \citealt{art-Iocco2008}; \citealt*{art-FreeseSpolyarAguirre2008}; \citealt{art-Taosoetal2008}; \citealt*{art-SchleicherBK2008}; \citealt{art-Natarajan2009,art-RipamontiIocooetal2010MNRAS,art-SivertssonGondolo2010}), including the prospects for their detection with the \textit{JWST} telescope \citep{art-FreeseISVB2010ApJ,art-ZackScottIoccoetal2010ApJ}. Similarly, other authors focused on the DM effects on stars in the local Universe, either on compact stars (\citealt{art-MoskalenkoWai2007,art-BertoneFairbairn2008,art-Isernetal2008ApJL,art-Isernetal2010A&A,art-LavallazFairbairn2010PhRvD,art-KouvarisTinyakov2010}; \citealt*{art-Perez-Garciaetal2010}) or on low-mass stars (\citealt*{art-Fairbairnetal2008,art-Scottetal2009MNRAS,art-CasanellasLopes2009}).

The purpose of this paper is to pave the way for the use of asteroseismology to provide an evidence of the footprint left by DM annihilation on the stellar oscillations. To do this, we will concentrate on solar-mass stars that evolve in haloes with very high DM densities, and we will show how asteroseismology may tell us about the properties of such DM haloes.

\section[]{Stellar evolution within dense dark matter haloes}
\label{sec-stellarevolution}
The evolution of a star within a halo of DM depends strongly on the ability of the gravitational field of the star to capture the DM particles that populate the halo. The rate at which the DM particles are captured is given by \citep{art-Gould1987}
\begin{equation}
    C_{\chi}(t) = \sum_i \int^{R_\star}_0 4\pi r^2\int^\infty_0 \frac{f_{v_{\star}}(u)}{u}w\Omega_{v_i}^-(w)\,\mathrm{d}u\,\mathrm{d}r\;,
\label{eq-cap}
\end{equation}
where $f_{v_{\star}}(u)$ is the velocity distribution of the DM particles seen by the star (which is proportional to the density of DM on the host halo $\rho_{\chi}$ and inversely proportional to the mass of the DM particles $m_{\chi}$) and $\Omega_{v_i}$ is the probability of a DM particle to be captured after the collision with an element $i$ (which is proportional to the scattering cross-section of the DM particle with the nucleus $i$, $\sigma_{\chi,i}$). The numerical subroutines to calculate the capture rate (equation \ref{eq-cap}) were adapted from the publicly available \begin{footnotesize}DARKSUSY\end{footnotesize} code \citep{art-GondoloEdsjoDarkSusy2004}. Our assumptions regarding this calculation are described in \cite{art-CasanellasLopes2010procIU}.

Once DM particles are captured, they accumulate in a small region in the core of the star ($r_{\chi}\simeq0.01\;$R$_{\star}$ for $m_{\chi}=100\;$GeV). There, assuming that they are Majorana particles, they annihilate providing a new source of energy for the star. Capture and annihilation processes balance each other in a short time-scale, and consequently almost all captured particles will be converted to energy, contributing to the total luminosity with $L_{\chi}=f_{\chi}C_{\chi}m_{\chi}$. The factor $f_{\chi}$, which in this work we assumed to be 2/3 \citep{art-FreeseBSG08,art-Ioccoetal2008,art-YoonIoccoAkiyama2008}, accounts for the energy that escapes out of the star in the form of neutrinos. Recent Monte Carlo simulations suggest that the fraction of the energy lost in neutrinos may be even smaller \citep{art-Scottetal2009MNRAS}.

Due to this new source of energy, stars will evolve differently from the classical picture if surrounded by a dense halo of DM. For very high DM densities ($\rho_{\chi}>3\times10^{9}\;$GeV$\;$cm$^{-3}$ for a 1 M$_{\odot}$ star), the energy from DM annihilation prevents the gravitational collapse of the star, stopping its evolution in the pre-main-sequence phase, before the star could reach enough central temperature to trigger hydrogen burning \citep{art-CasanellasLopes2009}.

For lower DM densities ($10^{8}\;$GeV$\;$cm$^{-3}<\rho_{\chi}<3\times10^{9}\;$GeV$\;$cm$^{-3}$ for a 1 M$_{\odot}$ star), DM burning is a complementary source of energy for the star. As it is produced in a region much more concentrated than the nuclear burning, which normally extends up to $0.1-0.2\;$R$_{\star}$, the radiative temperature gradient ($\bigtriangledown_{rad}=d\ln T / d\ln P_g$) is much steeper in the core of the star. Consequently, as the radiative transport is not efficient enough to evacuate all the energy in the central region, the star develops a convective core which was not present in the classical scenario without DM. The radius and duration of the convective core increase when more energy from DM annihilation is produced \citep{art-Scottetal2009MNRAS}; therefore, they depend on the density of DM in the place where the star evolves and on the properties of the DM particles. The balance between DM annihilation, nuclear burning, and the gravitational energy leads to a new hydrostatic equilibrium with a lower central temperature. The star consumes its hydrogen at a lower rate, extending the time that it spends in the main sequence. These new properties allow us, as it will be shown, to provide a tool to infer the DM characteristics from the stellar oscillations using the seismological analysis.

\section{Basics of asteroseismology}
\label{sec-asteroseismology}

With the improvement on the quality of the data, asteroseismology is now becoming a precise tool to infer the properties of stars showing solar-like oscillations \citep{art-CoRoT2008Sci,art-Garciaetal2009A&A,art-RGKepler2010ApJ}, which are driven by turbulence in the superficial layers of the star. The eigenfrequencies of solar-like oscillations can be approximated, for $l/n\rightarrow0$ (where $l$ and $n$ are the degree and the radial order of the modes), by the asymptotic expression
\begin{equation}
 \nu_{n,l}\simeq(n+\frac{l}{2×}+\epsilon_{\nu})\nu_0\; + O(\nu^{-2}),
\label{eq-assymp}
\end{equation}
where $\nu_0=[2\int_0^R \frac{dr}{c}]^{-1}$ is the inverse of twice the time spent by the sound to travel between the centre and the acoustic surface of the star, and $\epsilon_{\nu}$ is determined by the properties of the surface layers. For a more in-depth explanation of the basics of the seismological analysis, the reader is referred to \cite{art-Tassoul1980ApJS,art-Gough1985Natur,art-LopesTurck1994A&A290}. The value of $\nu_0$ can be estimated through the \textit{large separation} $\Delta\nu_{n,l}$:
\begin{equation}
 \Delta\nu_{n,l}=\nu_{n,l}-\nu_{n-1,l}\simeq \nu_0.
\label{eq-Delta}
\end{equation}
This parameter is sensitive to the mean density of the star: $\Delta\nu_{n,l}\propto(M/R^3)^{1/2}$ \citep{book-Cox1980}, while the \textit{small separation} $\delta\nu_{n,l}$, given by
\begin{equation}
 \delta\nu_{n,l}=\nu_{n,l}-\nu_{n-1,l+2},
\end{equation}
 is sensitive to the temperature and chemical gradient in the deep interior. 

In the last years, other relations between the frequencies of the oscillation modes were proposed (for a recent review, see \cite{art-Ch-DalsHoudek2009Ap&SS} or \cite{book-AertsCh-DaKurtz2010}), broadening the diagnostic potential of seismology. Among the possible diagnostic methods of convective cores and envelopes \citep{art-Monteiroetal1994A&A,art-Lopesetal1997ApJ,art-LopesGough2001MNRAS}, we highlight the ratios between the small separations and the large separations developed by \cite{art-RoxburghVorontsov2003A&A} in order to suppress the effects of the modelling of the near-surface layers:
\begin{equation}
r_{01}=\frac{d_{01}}{\Delta\nu_{n,1}}, \hspace{1cm}
r_{10}=\frac{d_{10}}{\Delta\nu_{n+1,0}},
\label{eq-ratios}
\end{equation}
where
\begin{equation}
d_{01}=\frac{1}{8}(\nu_{n-1,0}-4\nu_{n-1,1}+6\nu_{n,0}-4\nu_{n,1}+\nu_{n+1,0}),
\label{eq-d01}
\end{equation}
\begin{equation}
d_{10}=-\frac{1}{8}(\nu_{n-1,1}-4\nu_{n,0}+6\nu_{n,1}-4\nu_{n+1,0}+\nu_{n+1,1}).
\label{eq-d10}
\end{equation}

The mixing of elements produced in the convective regions introduces a sharp structural variation in the border with the radiative regions that can be seen in the density and sound-speed profiles. This sharp structural variation produces an oscillatory signal in the frequency spectrum \citep{art-Gough1990LNP} whose period is related with the acoustic depth of the discontinuity inside the star. Recently, \cite{art-Silvaetal2010} proposed the use of the ratios $r_{01}$ and $r_{10}$ to determine the size of a convective core by fitting a sine wave to their oscillation pattern. Similarly, another combination of the small and large separations,
\begin{equation}
 dr_{0213}\equiv \frac{D_{02}}{\Delta\nu_{n-1,1}}-\frac{D_{13}}{\Delta\nu_{n,0}},
\label{eq-CM}
\end{equation} where $D_{l,l+2}\equiv\delta\nu_{n,l}/(4l+6)$, was suggested by \cite{art-CunhaMetcalfe2007ApJ} to measure the amplitude of the sound-speed discontinuity at the edge of a convective core. These seismic parameters (equations \ref{eq-ratios} and \ref{eq-CM}) are sensitive to the presence of DM inside a star, given that they are uniquely dependent on the star's core structure and almost independent of the physical processes occurring in the surface layers.

\section[]{Asteroseismic signature of dark matter particles}
\label{sec-asterosesismDM}
\begin{table*}
\centering
\begin{minipage}{119mm}
\centering
\caption[]{Characteristics of stars of 1 M$_{\odot}$ when they reached a luminosity $L=1\;$L$_{\odot}$ after evolving in haloes of DM with different densities $\rho_{\chi}$ and different SD WIMP-nucleon cross-sections $\sigma_{\chi,SD}$. The last two columns are the radius and the acoustic radius ($\tau=\int_0^r \frac{dr}{c}$) of the convective core (CC). All the stars had the same initial conditions ($Z=0.018$).}
 \begin{tabular}[]{l c|c c c c c c c}
\hline
&$\rho_{\chi}\,\sigma_{\chi}$ & \multirow{2}{*}{$X_{c}$} & $R_{\star}$ & $T_{eff}$ & T$_c$ & $\rho_c$ & $r_{CC}$ & $\tau_{CC}$ \\
&(GeV cm$^{-1}$) & & (R$_{\odot}$) & (K) & (MK) & (g cm$^{-3}$) & (R$_{\star}$) & (s)\\ 
\hline
(i)& $0$ & 0.35 & 1.000 & 5777.5 & 15.52 & 148.7 & No CC & -\\
(ii)& $10^{-30}$ & 0.38 & 1.003 & 5768.1 & 15.66 & 137.0 & 0.04 & 53.6 \\
(iii)& $10^{-29}$ & 0.40 & 1.024 & 5708.5 & 15.25 & 123.8 & 0.08 & 110.8 \\
(iv)& $2\times10^{-29}$ & 0.38 & 1.047 & 5646.0 & 15.78 & 114.9 & 0.09 & 132.7 \\
(v)& $3\times10^{-29}$ & 0.35 & 1.071 & 5582.4 & 15.65 & 108.5 & 0.10 & 145.6 \\
\hline
 \end{tabular}
\label{tab-stars}
\end{minipage}
\end{table*}

\begin{figure}
\centering
 \includegraphics[scale=0.7]{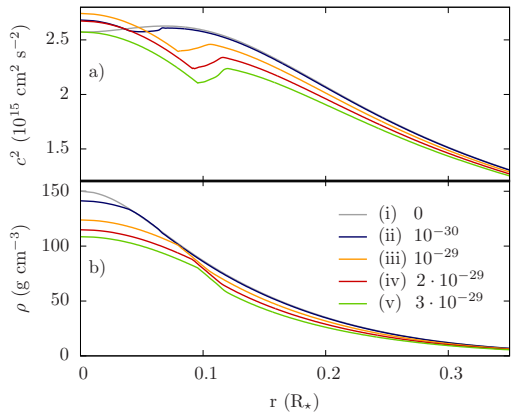}
\caption{Sound-speed (a) and density profiles (b) of 1 M$_{\odot}$ stars that evolved in DM haloes with different densities $\rho_{\chi}$  and SD WIMP-nucleon cross-sections $\sigma_{\chi,SD}$ when they reached a luminosity $L=1\;$L$_{\odot}$ (for each star, the product $\rho_{\chi}\sigma_{\chi}$ is indicated in the legend in GeV cm$^{-1}$).}
\label{fig-c2}
\end{figure}

To grasp the signature that the annihilation of captured DM particles leaves on low-mass stars we evolved a set of 1 M$_{\odot}$ stars, with the same initial conditions (Z=0.018), in haloes of DM with different densities $\rho_{\chi}$ and different spin-dependent (SD) WIMP-nucleon cross-sections $\sigma_{\chi,SD}$. Throughout our work, we considered fiducial values for the mass of the DM particles, $m_{\chi}=100\;$GeV, and for their self-annihilation cross-section, $<\sigma_a v>=3\cdot10^{-26}\;$cm$^3\;$s$^{-1}$. The evolution of the stars was computed using a well-established stellar evolution code (\begin{footnotesize}CESAM\end{footnotesize}; \cite{art-Morel1997}) used to compute sophisticated solar models for helioseismology \citep{art-CouvidatTuChi2003,art-Turck-Chiezeetal2010ApJ} and more recently used in the context of asteroseismic studies \citep{art-Kervella2004,art-DeRidderetl2006A&A,art-Suarezetal2010}. When the stars reached a luminosity equal to that of the Sun, a very precise mesh (with 1000 layers) was generated. Then, we calculated the frequencies of the oscillation modes of the stars using the \begin{footnotesize}ADIPLS\end{footnotesize} code \citep{art-Ch-Dals2008Ap&SS}. The characteristics of some of these stars are shown in Table \ref{tab-stars}, and their sound-speed and density profiles, in Fig. \ref{fig-c2}.

\begin{figure}
\centering
 \includegraphics[scale=0.7]{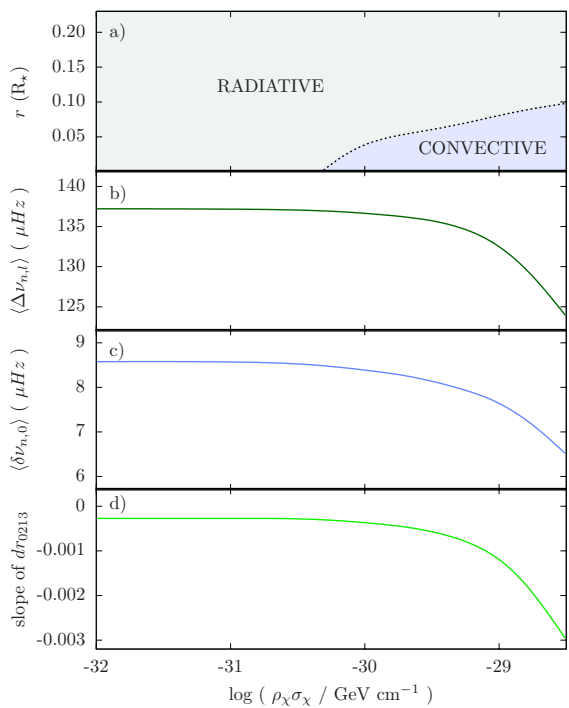}
\caption{(a) Size of the convective core, and the calculated seismologic parameters: (b) mean  large separation (for l=0,1,2,3), (c) mean small separation (for l=0) and (d) slope of $dr_{0213}$, for 1$\;$M$_{\odot}$ stars that evolved in DM haloes with different densities $\rho_{\chi}$ and SD WIMP-nucleon cross-sections $\sigma_{\chi,SD}$, when the stars reached a luminosity $L=1\;$L$_{\odot}$.}
\label{fig-multi}
\end{figure}

The accretion and the annihilation of DM particles in the core of the stars may change significantly their properties. As expected, we found that the effective temperature of the stars that evolved in haloes with high DM densities is shifted to lower values (see Table \ref{tab-stars}), due to the presence of a convective core (see Fig. \ref{fig-multi}.a), in agreement with previous works \citep{art-Fairbairnetal2008,art-CasanellasLopes2009}. The lower effective temperature and the larger radius lead to a decrease in the large separation $\Delta\nu_{n,l}$ (see Fig. \ref{fig-multi}.b). For a star with a known mass, the drop in $\Delta\nu_{n,l}$, predicted by the relation $\Delta\nu_{n,l}\propto M^{1/2}R^{-3/2}$, is unmistakably related with the radius of the star. Furthermore, we also observed a drop on the small separation $\delta\nu_{n,o}$ (see Fig. \ref{fig-multi}.c), caused by a decrease in the central density. The strong dependence of the global modes on the density profile of the star is responsible for that drop.

In order to test the validity of our method, we checked if classical stars with similar characteristics may mimic the properties we described for stars evolving in DM haloes. In particular, we found that a star with a mass $M_{\star}=0.955\;$M$_{\odot}$ and a metallicity $Z=0.04$ reaches, near the end of the main sequence, the same luminosity and effective temperature as the star (iv) in our set (see Table \ref{tab-stars}). At that moment, the radius of both stars is identical, leading to very similar great separations ($<\Delta\nu_{n,l}>=128\;\mu$Hz for star (iv) and $126\;\mu$Hz for the other). However, as the star that evolved without DM is in a later stage of evolution ($X_c=0.03$, while $X_c=0.38$ for star (iv), the small separation, being very sensitive to the chemical gradient in the deep interior, allows us to differentiate both stars. In our case, star (iv), which evolved in a dense halo of DM, has a $<\delta\nu_{n,o}>=7\;\mu$Hz. This is almost double than that of the star with different $M_{\star}$ and $Z$ ($<\delta\nu_{n,o}>=4\;\mu$Hz in that case).

In addition, one of the most promising signatures of annihilating DM in stars is the fact that it can originate the formation of a convective core (unexpected in the classical picture for stars with masses $<1.2$ M$_{\odot}$) whose radius grows with the DM density $\rho_{\chi}$. The convective core leaves a peculiar footprint in the profiles of the sound speed and density (see Fig. \ref{fig-c2}) characterized by a discontinuity in the edge of the core. The presence of the convective core can be detected by the seismological analysis using a relation between the small separation of modes with different degrees (and therefore with different depths of penetration inside the star). For that purpose low-degree modes ($l=0,1,2,3$) are chosen, because these modes are the ones that penetrate deep into the stellar core.

\begin{figure}
 \includegraphics[scale=0.7]{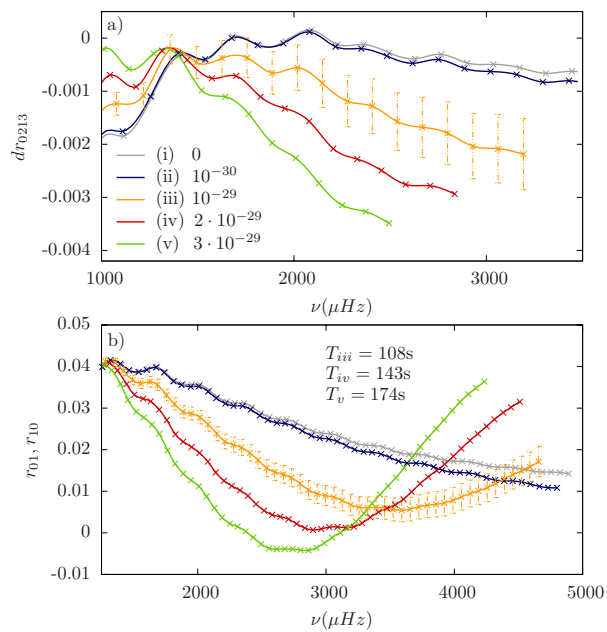}
\caption{(a) The seismological parameter $dr_{0213}$ and (b) the ratios $r_{01}$ and $r_{10}$ for stars that evolved in DM haloes with different densities $\rho_{\chi}$ and different SD WIMP-nucleon cross-sections $\sigma_{\chi,SD}$ when they reached a luminosity $L=1\;$L$_{\odot}$ (for each star, the product $\rho_{\chi}\sigma_{\chi}$ is indicated in the legend in GeV cm$^{-1}$). Error bars are shown for star (iii) assuming a relative error on the identification of the frequencies of $10^{-4}$.}
\label{fig-Delta_r01}
\end{figure}

In particular, we found that the seismological parameter $dr_{0213}$ (see equation \ref{eq-CM}) is sensitive to the sound-speed discontinuity at the edge of the convective core and, consequently, to the characteristics of the DM halo. In Fig. \ref{fig-Delta_r01}.a) we show the behaviour of the parameter $dr_{0213}$ for stars that evolve in DM haloes with different characteristics. We found that the absolute value of the slope of $dr_{0213}$ at high frequencies increases with the amplitude of the sound-speed discontinuity caused by the convective core, as predicted by Cunha and Metcalfe. Therefore, the slope of $dr_{0213}$ is directly related with the amount of DM in the halo where the star evolves (see also Fig. \ref{fig-multi}.d).

We also tested the method recently proposed by \cite{art-Silvaetal2010} designed to estimate the size of a convective core in 1.5 M$_{\odot}$ stars. We found that the period of the sinusoidal fit to the ratios $r_{01}$ and $r_{10}$ (see Fig. \ref{fig-Delta_r01}.b) does not match exactly the acoustic radius of the convective cores (see Table \ref{tab-stars}), most probably because we are applying this method to stars of mass 1 M$_{\odot}$. However, the ratios $r_{01}$ and $r_{10}$ have a great sensitivity to the amplitude of the sharp variation of the sound speed caused by the annihilation of DM particles inside the star. We conclude that these ratios may be used in the future as a stellar probe to confirm the presence of DM in the neighbourhood of low-mass stars. 

If enough radial modes are identified with the precision presently achieved by space-based telescopes as \textit{CoRoT} (a relative error on the individual frequencies of $\sim10^{-4}$ \citep{art-DeheuvelsetalCoRoT2010A&A515}), then our method will allow the discrimination between haloes of DM with different characteristics. To illustrate this point, we plotted in Fig. 3 the error bars on $dr_{0213}$, $r_{01}$ and $r_{10}$ for star (iii) derived from the mentioned uncertainty ($10^{-4}\nu$) on the determination of the frequencies, as done by \cite{art-CunhaMetcalfe2007ApJ}.

\section{Discussion and Conclusions}
\label{sec-Conclusions}

In this paper, we have presented a new methodology towards the use of asteroseismology to prove the presence of DM in the location where a star evolves. For a main-sequence star of 1 M$_{\odot}$ evolving in a DM halo with a density $\rho_{\chi}>10^{8}\;$GeV$\;$cm$^{-3}$ (assuming $\sigma_{\chi,SD}=10^{-38}\;$cm$^2$), the annihilation of captured DM particles on its interior leads to decreases in the large and small separations, when compared with the same star in the classical scenario without DM, which are related to changes in the global properties of the star. Furthermore, the highly concentrated production of energy by DM annihilation creates a convective core which is not present in the classical picture. This convective core leaves a discontinuity signature in the sound-speed and density profiles which can be detected by the analysis of the stellar oscillations.

We have shown that seismological parameters such as $dr_{0213}$ and the ratios $r_{01}$ and $r_{10}$ are very sensitive to the size of the convective core, which is determined by the density of DM $\rho_{\chi}$ where the star evolved and by the scattering cross-section of the DM particles off nuclei $\sigma_{\chi}$. Consequently, this relationship may be used in the future to help in the determination of these parameters (or at least to their product, $\rho_{\chi}\sigma_{\chi}$) and to provide a stellar probe that identifies the presence of self-annihilating DM.
\begin{figure}
\centering
\includegraphics[scale=0.4]{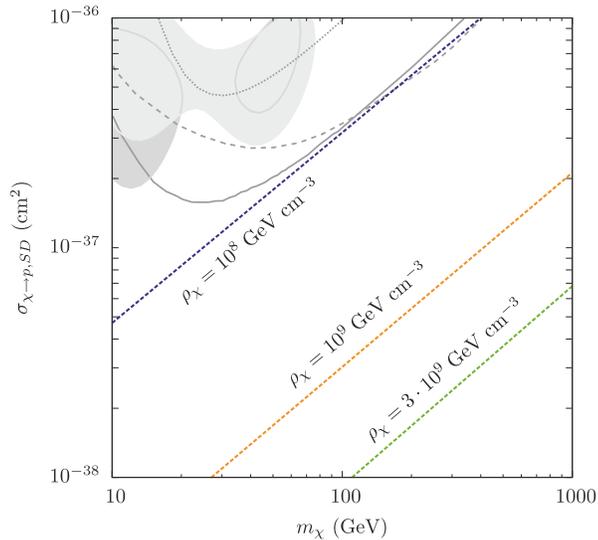}
\caption{DM densities at which 1 M$_{\odot}$ stars with a luminosity 1 L$_{\odot}$ are expected to show strong signatures on the seismological parameters $\Delta\nu, \delta\nu, dr_{0213}, r_{01}$ and $r_{10}$ (see text) due to the annihilation of DM particles with different characteristics ($m_{\chi},\sigma_{\chi,SD}$) in their interior. In the particular case of the Galactic Centre (GC), the DM densities in the Figure (from top to bottom) are expected at a distance from the GC of 0.1 pc, 0.04 pc and 0.02 pc, following the adiabatically contracted profile of Bertone \& Merritt. The grey lines are the present limits from direct detection experiments: XENON10 (dotted), PICASSO (dashed) and COUPP (solid), and the grey region is the DAMA/LIBRA allowed region.}
\label{fig-mxSD-rho}
\end{figure}

The method presented in this paper is valid for haloes with very high DM densities. In Fig. \ref{fig-mxSD-rho} we show the DM densities at which a 1 M$_{\odot}$ star with a luminosity 1 L$_{\odot}$ is expected to have a small separation 25\% smaller than that in the classical picture, because of the annihilation of DM particles with different characteristics ($m_{\chi},\sigma_{\chi\rightarrow p,SD}$) in its interior. If found, this kind of star will have also strong signatures on the seismic parameters $\Delta \nu$, $dr_{0213}$, $r_{01}$ and $r_{10}$ when compared with a star with the same luminosity that evolved without DM. In the same Figure are also shown the current limits on $\sigma_{\chi\rightarrow p,SD}$ from the direct detection experiments XENON10 \citep{art-XENON10_SD2008}, PICASSO \citep{art-PICASSOSD2009}, COUPP \citep{art-COUPP2008} and the allowed region from the DAMA/LIBRA experiment \citep{art-Savageetal2009JCAP}. 

The extreme DM densities shown in Figure \ref{fig-mxSD-rho} may be present within the inner parsec of our Galaxy, according to models that account for the effect of the baryons on the DM halo via adiabatic contraction \citep{art-Blumenthaletal1986ApJ301,art-Gnedinetal2004ApJ616}. For instance, following the adiabatically contracted profile of \cite{art-BertoneMerritt2005}, DM densities as high as $\rho_{\chi}=10^{8}\;$GeV$\;$cm$^{-3}$ are expected at 0.1 pc from the Galactic Centre (GC). Even higher DM densities may be present at the GC if a hypothetical spike is formed due to the influence of the central black hole \citep{art-GondoloSilk1999}. However, as other models predict lower central DM densities (the so-called \textit{core models} \citep{art-Burkert1995ApJ447}), the open questions about the DM halo profile at the inner parsec of our Galaxy are still far from being solved (for a recent review on this topic, see \cite{art-MerrittChap10Bertone} and \cite{art-deBlok2010AdAst}). In this sense, the method proposed here may provide a complementary tool to help in the discrimination of different models. Other possible locations of environments with such high DM densities are the dwarf spheroidal galaxies around the Milky Way \citep{art-DekelSilk1986ApJ,art-KormendyFreeman2004IAUS,art-Diemandetal2007ApJ}.

The precision required for our analysis is similar to the one achieved by present asteroseismic missions in observations of one hundred days. Nevertheless, the most likely place to find the kind of stars described here is near the centre of our Galaxy, where the distance and the presence of dust makes the observations difficult. These difficulties encourage us to extend our study to more massive and luminous stars in a future work. Future technical improvements in the observations of the GC and of the Milky Way dwarf galaxies may open the possibility of using the method proposed here to investigate the nature of DM.

\begin{flushleft}
{\bf Acknowledgements}
\end{flushleft}

We acknowledge the anonymous referee for his useful comments, as well as the authors of \begin{footnotesize}CESAM, ADIPLS\end{footnotesize} and \begin{footnotesize}DARKSUSY\end{footnotesize}, and the Brown University's Particle Astrophysics Group, which maintains the DM tools website, used for the $\sigma_{\chi\rightarrow p,SD}$ limits in Figure \ref{fig-mxSD-rho}. This work was supported by grants from ``Funda\c c\~ao para a Ci\^encia e Tecnologia" (SFRH/BD/44321/2008) and ``Funda\c c\~ao Calouste Gulbenkian".

\bibliography{DM}
\end{document}